\documentclass[prd,showpacs,showkeys,floatfix,
               12pt,preprint,
               tightenlines,fleqn,eqsecnum]{revtex4}

\usepackage{amsmath,amsfonts,latexsym,graphicx,mathrsfs}

\newcommand {\version}{v5} 

\newcommand{\st}{spacetime}
\newcommand{\dr}{dispersion relation}
\newcommand{\drs}{dispersion relations}
\newcommand{\beq}{\begin{equation}}
\newcommand{\eeq}{\end{equation}}
\newcommand{\beqa}{\begin{eqnarray}}
\newcommand{\eeqa}{\end{eqnarray}}
\newcommand{\no}{\nonumber}
\newcommand{\bR}{\mathbb{R}}
\newcommand{\bC}{\mathbb{C}}

\renewcommand{\i}{{\rm i}}

\newcommand{\lhs} {left-hand side}
\newcommand{\rhs} {right-hand side}
\newcommand{\bc}   {boundary condition}
\newcommand{\bcs}  {boundary conditions}

\begin{document}
\noindent Phys. Rev. D 75, 024028 (2007) \hfill
hep-ph/0610216 (\version)\vspace*{2\baselineskip}
\title{Bounds on length scales of classical spacetime foam models}

\author{S. Bernadotte}

\author{F.R.\ Klinkhamer}
\email{frans.klinkhamer@physik.uni-karlsruhe.de}
\affiliation{Institute for Theoretical Physics, University of Karlsruhe (TH),
            76128 Karlsruhe, Germany  \\ \\}

\begin{abstract}
Simple models of a classical spacetime foam  are considered,
which consist of identical static defects embedded in Minkowski spacetime.
Plane-wave solutions of the vacuum Maxwell equations with
appropriate boundary conditions at the defect surfaces are obtained in the
long-wavelength limit. The corresponding dispersion relations
$\omega^2=\omega^2(\vec{k})$  are calculated, in particular,
the coefficients of the quadratic and quartic terms in $\vec{k}$.
Astronomical observations of gamma--ray bursts and
ultra-high-energy cosmic rays then place bounds on the coefficients of
the dispersion relations and, thereby, on particular combinations of the
fundamental length scales of the static spacetime-foam models considered.
Spacetime foam models with a single length scale are excluded,
even models with a length scale close to the Planck length
(as long as a classical spacetime remains relevant).
\end{abstract}



\maketitle

\section{Introduction}
\label{sec:intro}

Whether or not space remains smooth down to smaller and smaller
distances is an open question.
Conservatively, one can say that the typical length scale of any
fine-scale structure of space must be less than $10^{-18}\,\text{m}
= 10^{-3} \,\text{fm}$,
which corresponds to the inverse of a center-of-mass energy of
$200\,\text{GeV}$ in a particle-collider experiment.
Astrophysics provides us, of course, with much higher energies,
but not with controllable experiments. Still, astrophysics
may supply valuable information as long as the relevant physics
is well understood.

In this article, we discuss astrophysical bounds
solely based on solutions of the Maxwell (and Dirac) equations.
These solutions hold for a particular
type of classical spacetime with nontrivial small-scale structure.
Specifically, we consider a static (time-independent)  fine-scale
structure of space, which is modeled by a homogeneous and
isotropic distribution of identical static  ``defects'' embedded
in Minkowski spacetime. With appropriate \bcs~at the defect surfaces,
plane-wave solutions of the vacuum Maxwell equations are obtained
in the long-wavelength limit.
That is, the wavelength $\lambda$ must be much larger than
$\max(b,l)$, with $b$ the typical size of the individual defect and $l$
the mean separation between the different defects.
An (imperfect) analogy would be sound propagation in a block
of ice with frozen-in bubbles of air.

Generalizing the terminology of Wheeler and
Hawking \cite{W57,W68,H78,HPP80,Fetal90,V96},
we call \emph{any} classical spacetime with nontrivial
small-scale structure (resembling bubbly ice, Swiss cheese, or whatever)
a ``classical spacetime foam.''
The plane-wave Maxwell solutions from our classical
spacetime-foam models, then,
have a modified dispersion relation (angular frequency squared as a
function of the wave number $k\equiv |\vec{k}| \equiv 2\pi/\lambda$):
\begin{widetext}
\beqa
\omega^{2}_\gamma \,\Big|^{[\text{defect type}\,\tau]}
&=&
a_{\gamma\,2}^{[\tau]}\;\, c^2\, k^2
+a_{\gamma\,4}^{[\tau]}\; \big(\, b^{[\tau]} \,\big)^2\, c^2\, k^4
+ \ldots\,,
\label{disprel-photon}
\eeqa
\end{widetext}
where $c$ is the characteristic velocity of the Minkowski line element
($\text{d}s^2 = c^2\,\text{d}t^2 -|\text{d}\vec x|^2$\,),
$a_{\gamma\,2}$ and $a_{\gamma\,4}$ are dimensionless coefficients
depending on the fundamental length scales of the model
(one length scale being $b$), and $\tau$ labels different kinds of models.

For simplicity, we consider only three types of static defects
(or ``weaving errors'' in the fabric of space):
\begin{enumerate}
\item[(i)]
a nearly pointlike defect with the interior of a ball removed
from $\mathbb{R}^3$ and antipodal points on its boundary identified;
\item[(ii)]
a nearly pointlike defect with the interior of a ball removed
from $\mathbb{R}^3$ and
boundary points reflected in an equatorial plane identified;
\item[(iii)]
a wormhole-like defect with two balls removed from $\mathbb{R}^3$ and glued
together on their boundaries;  cf. Refs.~\cite{Fetal90,V96}.
\end{enumerate}
Further details will be given in Sec.~\ref{sec:defect-types}.
As mentioned above, the particular spacetime models considered consist
of a frozen gas of \emph{identical} defects (types $\tau=1$, $2$, $3$)
distributed homogeneously and isotropically over
Euclidean space $\mathbb{R}^3$. We emphasize that these classical models
are not intended to describe in any detail a possible
spacetime-foam structure (which is, most likely, essentially
quantum-mechanical in nature) but are meant to provide
simple and clean backgrounds for explicit calculations of
potential nonstandard propagation effects of electromagnetic waves.

The type of Maxwell solution found here is reminiscent
of the solution from the so-called ``Bethe holes'' for waveguides
\cite{B44}. In both cases, the standard Maxwell plane wave is modified by
the radiation from fictitious multipoles
located in the holes or defects. But there is a crucial difference:
for Bethe, the holes are in a material conductor, whereas for us,
the defects are ``holes'' in space itself.

Returning to our spacetime-foam models, we also
calculate the modified dispersion relation of
a free Dirac particle, for definiteness taken to be a proton
(mass $m_p$):
\begin{widetext}
\beq
\omega^{2}_p \,\Big|^{[\text{defect type}\,\tau]}=
a_{p\,0}^{[\tau]}\;\, \hbar^{-2}\,c^4\, m_p^2 +
a_{p\,2}^{[\tau]}\;\, c^2\, k^2 +
a_{p\,4}^{[\tau]}\;  \big(\, b^{[\tau]} \,\big)^2\, c^2\, k^4 + \ldots\,,
\label{disprel-Dirac}
\eeq
\end{widetext}
with reduced Planck constant $\hbar \equiv h/2\pi$
and dimensionless coefficients $a_{p\,0}$, $a_{p\,2}$,
and $a_{p\,4}\,$. As might be expected, the response of
Dirac and Maxwell plane waves to the same spacetime-foam model turns
out to be quite different, with unequal quadratic
coefficients $a_{p\,2}$ and $a_{\gamma\,2}$, for example.
The different proton and photon velocities then allow for
Cherenkov--type processes \cite{Beall1970,ColemanGlashow1997}.
But, also in the \emph{pure} photon sector, there can be interesting
time-dispersion effects \cite{AEMNS98} as long as the
quartic coefficient $a_{\gamma\,4}$ of the
photon \dr~\eqref{disprel-photon} is nonvanishing.

In fact, with the model dispersion relations in place, we may
use astronomical observations to put bounds on the various
coefficients $a_2$ and $a_4$, and, hence, on particular
combinations of the model length scales
(e.g., average defect size $\overline{b}$
and separation $\overline{l}\,$).
Specifically, the absence of time dispersion in an observed $\text{TeV}$
flare from an active galactic nucleus
bounds $|a_{\gamma\,4}|$ and the absence of Cherenkov--like effects in
ultra-high-energy cosmic rays
bounds $(a_{\gamma\,2}-a_{p\,2})$ and $a_{\gamma\,4}$.
In other words, astrophysics not only explores the largest
structures of space (up to the size of the visible universe
at approximately $10^{10}\,\text{lyr}  \approx 10^{26}\,\text{m}$)
but also the smallest structures (down to $10^{-26}\,\text{m}$ or less,
as will be shown later on).

The outline of the remainder of this article is as follows.
In Sec.~\ref{sec:calculation}, we discuss
the calculation of the effective photon dispersion relation
from the simplest type of foam model, with static $\tau=1$ defects.
The calculations for isotropic $\tau=2$ and $\tau=3$ models
are similar and are not discussed in detail
(App.~\ref{sec:birefringence} gives  additional results for
anisotropic defect distributions).
Some indications are, however, given for
the calculation of the proton dispersion relation
from model $\tau=1$ with  details relegated to
App.~\ref{sec:Dirac-wave-function}.
The main focus of Sec.~\ref{sec:calculation} and the two appendices
is on modified dispersion relations but in Sec.~\ref{sec:scattering}
we also discuss the Rayleigh-like scattering of an incoming
electromagnetic wave by the model defects.
In Sec.~\ref{sec:results}, we summarize the different
dispersion relations calculated and put the results in a general
form. In Sec.~\ref{sec:astrophysics-bounds}, this general photon
dispersion relation is confronted to the astronomical
observations and bounds on the effective length scales are obtained.
In Sec.~\ref{sec:conclusion}, we draw an important conclusion
for the classical small-scale structure of space and present some
speculations on a hypothetical quantum spacetime foam.

\section{Calculation}
\label{sec:calculation}

\subsection{Defect types}
\label{sec:defect-types}

The present article considers three types of static defects
obtained by surgery on the Euclidean 3--space
$\bR^3$. The discussion is simplified by initially choosing the
origin of the Cartesian coordinates
$\vec{x} \equiv (x^1, x^2, x^3) \equiv (x,y,z)$
of $\mathbb{R}^3$ to coincide with the ``center'' of the defect.
The corresponding Minkowski spacetime $\bR \times\bR^3$ has standard
metric $(\eta_{\mu\nu}) = \text{diag}(1,-1,-1,-1)$
for coordinates $x^\mu=(x^0,\vec{x})=(c\,t,x^m)$
with index $\mu=0$, $1$, $2$, $3$.

\begin{figure}[h] \vspace*{1cm} 
\begin{center}
\includegraphics[width=6cm]{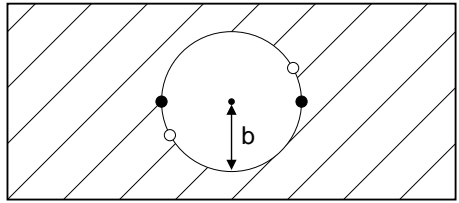}
\caption{\label{fig:defecttau1}
Three-space
\eqref{Mtau1} from a single
spherical defect (type $\tau=1$, radius $b$) embedded
in $\bR^3$, with its ``interior'' removed
and antipodal points identified (as indicated by
the pairs of open and filled circles).
The corresponding classical spacetime-foam model has a  homogeneous
distribution of static $\tau=1$ defects embedded in Minkowski spacetime.}
\end{center}
\end{figure}

The first type of defect (label $\tau = 1$) is obtained by
removing the interior of a ball from $\mathbb{R}^3$ and
identifying antipodal points on its boundary.
Denote this ball, its boundary sphere, and point reflection by
\begin{subequations}
\beqa
B_b &=& \big\{ \vec{x} \in \mathbb{R}^3 :\, |\vec{x}| \leq b \big\}\,,\quad
\\[1mm]
S_b &=& \big\{ \vec{x} \in \mathbb{R}^3 :\, |\vec{x}| = b \big\}\,,\quad
\\[1mm]
P(\vec{x}) &=& -\vec{x}\,.
\eeqa
\end{subequations}
Then, the 3--space with a single defect
centered at the origin $\vec{x}=0$ is given by
\begin{widetext}
\beq
\mathsf{M}^{[\tau = 1]}_{b} \equiv \mathsf{M}^{[\tau = 1]}_{0,\,b} =
\big\{ \vec{x} \in \mathbb{R}^3 :\, |\vec{x}| \geq b  \,\wedge\,
\left(  S_b \ni \vec{x} \;\cong\;   P(\vec{x}) \in S_b\right)\big\}\,,
\label{Mtau1}
\eeq
\end{widetext}
where $\cong$ denotes pointwise identification
and $\mathsf{M}^{[\tau = 1]}_{b}$ is a
shorthand notation. The 3--space \eqref{Mtau1} has no boundary
because of the $S_b$
identifications (cf. Figure \ref{fig:defecttau1}) and, away
from the defect at $|\vec x|=b$, is certainly a manifold
(hence, the suggestive notation $\mathsf{M}$). The resulting spacetime
is $\mathcal{M} = \mathbb{R} \times \mathsf{M}^{[\tau=1]}_b\,$.

The corresponding classical spacetime-foam model
is obtained from a superposition
of $\tau=1$ defects with a homogeneous distribution. The number
density of defects is denoted $n \equiv l^{-3}$ and only the case of a
very rarefied gas of defects is considered ($b \ll l$), so that
there is no overlap of defects. Clearly, there is a preferred
reference frame for which the defects are static.
Such a preferred frame, in the context of cosmology, may or may not
be related to the preferred frame of the isotropic
cosmic microwave background.

In more detail, the construction is as follows.
The 3--space with $N \geq 1$ identical defects is given by
\beq
\mathsf{M}^{[\tau = 1]}_{\{ \vec{x}_1,\,\ldots,\,\vec{x}_N \},\,b} =
\mathsf{M}^{[\tau = 1]}_{\vec{x}_1,\,b} \cap
\mathsf{M}^{[\tau = 1]}_{\vec{x}_2,\,b} \cap
\ldots \cap
\mathsf{M}^{[\tau = 1]}_{\vec{x}_N,\,b}\;,
\label{Mtau1-many}
\eeq
where $\mathsf{M}^{[\tau = 1]}_{\vec{x}_n,\,b}$ is the single-defect
3--space \eqref{Mtau1} with the center of the sphere
moved from $\vec{x} = 0$ to $\vec{x} = \vec{x}_n$.
The minimum distance between the different centers $\vec{x}_n$ of
3--space \eqref{Mtau1-many}
is assumed to be larger than $2 b$. The final spacetime foam model
results from taking the Cartesian product of
$\bR$ with the $N \to \infty$ ``limit'' of \eqref{Mtau1-many},
\beq
\mathcal{M}^{[\tau = 1]}_{\text{distribution},\,b}
= \bR \times \left( \,\lim_{N \to \infty}
\mathsf{M}^{[\tau = 1]}_{\{ \vec{x}_1,\,\ldots,\,\vec{x}_N \},\,b}\,\right)\,,
\label{Mtau1-foam}
\eeq
where one needs to give the statistical distribution of
the centers $\vec{x}_n$. As mentioned above, we choose
the simplest possible distribution, homogeneous,
and the quantity to specify is the number density $n$ of defects.

\begin{figure}[h] \vspace*{-.5cm}  
\begin{center}
\includegraphics[width=6cm]{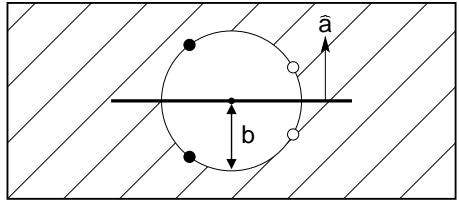}
\caption{\label{fig:defecttau2}
Three-space
\eqref{Mtau2} from a single
spherical defect (type $\tau=2$, radius $b$) embedded
in $\bR^3$, with its ``interior'' removed
and points identified by reflection in the equatorial plane with
normal vector $\widehat{a}$.
The corresponding classical spacetime-foam model has a  homogeneous
and isotropic distribution of static $\tau=2$ defects
embedded in Minkowski spacetime.}
\end{center}
\end{figure}

The second type of defect ($\tau=2$) follows by the same
construction, except that the identified points of the sphere
$S_b$ are obtained by reflection in an equatorial plane with
unit normal vector $\widehat{a}$.
For a point $\vec{x}$ on the sphere
$S_b$, the reflected point is denoted $R_{\widehat{a}}(\vec{x})$.
[With only one defect present, global Cartesian coordinates can be chosen
so that $\widehat{a}$ points in the 3--direction and
$(x^1, x^2, x^3)\in S_b$ is to be identified with $(x^1, x^2, -x^3)\in S_b$.]
The space with a single $\tau=2$ defect centered at the
origin $\vec{x}=0$ (not indicated by our shorthand notation)
is then given by  (cf. Figure \ref{fig:defecttau2})
\begin{widetext}
\beq
\mathsf{M}^{[\tau=2]}_{\widehat{a},\,b} =
\big\{ \vec{x} \in \mathbb{R}^3 :\, |\vec{x}| \geq b  \,\wedge\,
\left( S_b \ni \vec{x} \;\cong\;   R_{\widehat{a}}(\vec{x}) \in S_b\right)\big\}\,,
\label{Mtau2}
\eeq
\end{widetext}
and spacetime is $\mathbb{R} \times \mathsf{M}^{[\tau=2]}_{\widehat{a},\,b}$.
However, the defect embedded in \eqref{Mtau2} is not a manifold but an
orbifold \cite{Nakahara90}, i.e., a coset space $\mathsf{M}/G$, for
manifold $\mathsf{M}$ and discrete symmetry group $G$.
The 3--space \eqref{Mtau2} has, in fact, singular points corresponding
to the fixed points of $R_{\widehat{a}}$, which lie on the great circle
of $S_b$ in the equatorial plane with normal vector $\widehat{a}$.
But away from these singular points, the 3--space
is a genuine manifold and we simply use the (slightly misleading)
notation $\mathsf{M}$ in \eqref{Mtau2}.
The corresponding classical spacetime-foam model results from a homogeneous
and isotropic (randomly oriented) distribution of $\tau=2$ defects.

\begin{figure}[h] \vspace*{-.5cm}   
\begin{center}
\includegraphics[width=6cm]{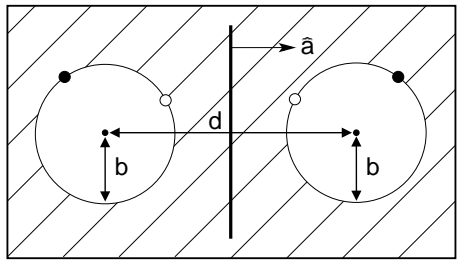}
\caption{\label{fig:defecttau3}
Three-space
\eqref{Mtau3} from a single
wormhole-like defect (type $\tau=3$, two spheres with
radii $b$ and distance $d$ between the centers) embedded
in $\bR^3$, with the ``interiors'' of the two spheres removed
and their points identified by reflection in the central plane with
normal vector $\widehat{a}$.
The corresponding classical spacetime-foam model has a  homogeneous
and isotropic distribution of static $\tau=3$ defects
embedded in Minkowski spacetime.}
\end{center}
\end{figure}

The third type of defect ($\tau=3$) is obtained by a somewhat more
extensive surgery \cite{Fetal90,V96}.
Now, the interiors of \emph{two} identical balls
are removed from $\mathbb{R}^3$. These balls,  denoted $B_b$ and
$B_b^\prime$, have their centers separated by a distance $d > 2 b$.
The two boundary spheres $S_b$ and $S_b^\prime$ are then pointwise
identified by reflection in the central plane.
This reflection is again denoted $R_{\widehat{a}}$. [With ball centers at
$\vec{x}=(\pm d/2,0,0)$, the reflection plane is given by $x^1=0$.]
The space manifold with a single $\tau=3$ defect centered at
the origin $\vec{x}=0$ is now  (cf. Figure \ref{fig:defecttau3})
\begin{widetext}
\beq
\mathsf{M}^{[\tau=3]}_{\widehat{a},b,d} =
\big\{ \vec{x} \in \mathbb{R}^3 :\,
|\vec{x}-(d/2)\,\widehat{a}\,| \geq b  \,\wedge\,
|\vec{x}+(d/2)\,\widehat{a}\,| \geq b  \,\wedge\,
\left( S_b \ni \vec{x} \;\cong\;
R_{\widehat{a}}(\vec{x}) \in S_b^\prime\,\right)\big\}\,,
\label{Mtau3}
\eeq
\end{widetext}
and the spacetime manifold is
$\mathbb{R} \times \mathsf{M}^{[\tau=3]}_{\widehat{a},b,d}$.
Using standard wormhole terminology \cite{W57,W68,H78,HPP80,Fetal90,V96},
the static $\tau=3$ defect has two ``wormhole mouths'' of diameter $2 b$,
with corresponding points on the wormhole mouths
separated by a vanishing distance through the ``wormhole throat''
and by a ``long distance'' $D \in [d-2b,d+\pi b]$
via the ambient Euclidean space.
Again, the corresponding classical spacetime-foam model results from a
homogeneous and isotropic distribution of $\tau=3$ defects.

For later use, we also define a $\tau=3^\prime$ defect with distance
$d$ set to the value $4 b$, where the factor $4$ has been chosen arbitrarily.
An individual  $\tau=3^\prime$ defect then has only one length scale, $b$,
which simplifies some of the discussion later on.
The relevant space manifold is thus
\beq
\mathsf{M}^{[\tau=3^\prime]}_{\widehat{a},\,b}\equiv
\mathsf{M}^{[\tau=3]}_{\widehat{a},b,4 b} \;,
\label{Mtau3prime}
\eeq
in terms of the $\tau=3$ manifold defined by \eqref{Mtau3}.

Let us end this subsection with two parenthetical remarks,
one mathematical and one physical.
First, the $\tau=1$ and $\tau=3$ spaces are multiply connected
(i.e., have noncontractible loops) but not the $\tau=2$ space.
Second, the classical spacetimes considered in this article
do not solve the vacuum Einstein equations but appear to require some
exotic form of energy located at the defects;
see, e.g., Part~III of Ref.~\cite{V96} for further discussion.
In one way or another, the fine-scale structure of spacetime may
very well be related to the ``dark energy'' of cosmology;
see, e.g. Refs.~\cite{PeeblesRatra2002,Spergel-etal2006} and
references therein.


\subsection{Photon dispersion relation}
\label{sec:photon-dispersion-relation}

The task, now, is to determine the electromagnetic-wave
properties for the three types of classical spacetime-foam models considered.
The $\tau=1$ case will be discussed in some detail
but not the other cases ($\tau=2$, $3$), for which only results will be given.

The calculation is relatively straightforward and consists of three steps.
First, recall the vacuum Maxwell equations
in Gaussian units \cite{PP62,F64,J75},
\begin{subequations}
\beqa
\nabla\cdot\vec{E} &=& 0\,,\quad c\,\nabla\times\vec{E}+\partial_t\vec{B}=0\,,
\\[1mm]
\nabla\cdot\vec{B} &=& 0\,,\quad c\,\nabla\times\vec{B}-\partial_t\vec{E}=0\,,
\label{vacuum-Maxwell-equations}
\eeqa
\end{subequations}
and the standard plane-wave solution over Minkowski spacetime,
\begin{subequations}\label{standard-plane-wave}
\beqa
\vec E_0(\vec{x},t) &=& E_0\,\widehat{x}\;\exp(\i\, k z -\i\, \omega_\gamma t)\,,
\label{standard-plane-wave-Efield}
\\[1mm]
\vec B_0(\vec{x},t) &=& E_0\,\widehat{y}\;\exp(\i\, k z -\i\, \omega_\gamma t)\,,
\label{standard-plane-wave-Bfield}
\eeqa
\end{subequations}
with amplitude $E_0$ and \dr~$\omega_\gamma^2 = c^2\,k^2$. This particular
solution corresponds to a linearly polarized plane wave  propagating
in the $z\equiv x^3$ direction ($\widehat{x}$ and $\widehat{y}$ are
unit vectors pointing in the $x^1$ and $x^2$ directions, respectively).
The pertinent  observation, now, is that
the electromagnetic fields \eqref{standard-plane-wave} also provide
a valid solution of the Maxwell equations between the ``holes'' of
the classical spacetime-foam models considered,
for example, model \eqref{Mtau1-foam} for $\tau=1$ defects.

Second, add appropriate vacuum solutions ($\vec E_1, \vec B_1$), so that
the total electric and magnetic fields,
$\vec E=\vec E_0+\vec E_1$ and
$\vec B=\vec B_0+\vec B_1$,
satisfy the \bcs~from a single defect.
The specific \bcs~for the electric field at the defect surface follow by considering
the allowed motions of an electrically charged  test particle
and similarly for the \bcs~of the magnetic field.
Geometrically, the electromagnetic-field \bcs~trace back to the
proper identification of the defect surface points and their tangential spaces.

Third, sum over the contributions ($\vec E_j, \vec B_j$) of the
different defects ($j=1$, $2$, $3$, $\ldots$)
in the model spacetime foam and obtain the effective dielectric and
magnetic permeabilities, $\epsilon$ and $\mu$, which may be wavelength
dependent. The dispersion relation for the isotropic case is then given by
\beq
\omega_\gamma^2 (k) = c^2\,k^2/\big(\epsilon(k)\,\mu(k)\big)\,,
\label{disprel}
\eeq
and we refer the reader to the textbooks for further discussion
(see, e.g., Sec.~II--32--3 of Ref.~\cite{F64} and
Secs.~7.5(a)  and 9.5(d)  of Ref.~\cite{J75}).

The specifics of the second and third step of the calculation
for $\tau=1$ defects are as follows. In step 2,
the motion of a test particle under influence of the Lorentz force
(see, for example, the points marked in Fig.~\ref{fig:defecttau1}
for tangential motion) gives the following \bcs~at the surface $S_b$ of
3--space \eqref{Mtau1}:
\begin{widetext}
\begin{subequations}
\label{EBbcs}
\beqa
\hspace*{-9mm}
 \widehat{n}(\vec{x})  \cdot \vec E(\vec{x},t) &=&
-\widehat{n}(-\vec{x}) \cdot \vec E(-\vec{x},t)\,,\;\;
 \widehat{n}(\vec{x})  \times\vec E(\vec{x},t)=
+\widehat{n}(-\vec{x}) \times \vec E(-\vec{x},t)\;
\Big|_{\,|\vec x|=b}^{\,[\tau = 1]}\,,
\label{Ebcs}\\[2mm]
\hspace*{-9mm}
 \widehat{n}(\vec{x})  \cdot \vec B(\vec{x},t)&=&
+\widehat{n}(-\vec{x}) \cdot \vec B(-\vec{x},t)\,,\;\;
 \widehat{n}(\vec{x})  \times\vec B(\vec{x},t)=
-\widehat{n}(-\vec{x}) \times \vec B(-\vec{x},t)\;
\Big|_{\,|\vec x|=b}^{\,[\tau = 1]}\,,
\label{Bbcs}
\eeqa
\end{subequations}
\end{widetext}
where $\widehat{n}(\vec{x})$ is the \emph{outward} unit normal vector
of the surface at point $\vec{x}\in S_b \subset \bR^3$.
The \bcs~\eqref{EBbcs} also ensure an equal Poynting vector
$c (\vec{E}\times\vec{B})/4\pi$ at antipodal points,
as might be expected for an energy-flux density passing through
the defect. As mentioned above, these \bcs~trace back to
the antipodal identification of the points on the surface $S_b$
and the identification of the respective tangential spaces.

Constant fields $\vec E \propto E_0\,\widehat{x}$
and $\vec B \propto E_0\,\widehat{y}$, corresponding
to the unperturbed fields \eqref{standard-plane-wave}
over distances of order $b \ll \lambda$,
do not satisfy the defect \bcs~\eqref{EBbcs}
and need to be corrected. As discussed in the Introduction,
the correction fields $\vec E_1$ and $\vec B_1$ correspond to multipole
fields from ``mirror charges'' located inside the defect.
The leading contributions of a $\tau=1$ defect come
from fictitious  electric and  magnetic dipoles at the defect center,
each aligned with their respective
initial fields \eqref{standard-plane-wave} and
both with a strength proportional to $b^3\,E_0$.
For $\tau=1$ and $k b \ll 1$, the electric field
$\vec E=\vec E_0+\vec E_1$
turns out to be normal to the surface $S_b$ and
the magnetic field $\vec B=\vec B_0+\vec B_1$ tangential,
just as for a perfectly conducting sphere
(see, e.g., Secs.~13.1 and 13.9 of Ref.~\cite{PP62}).

In step 3 of the calculation for $\tau=1$ defects,
the effective dielectric and magnetic permeabilities (Gaussian units)
are found to be given by
\begin{subequations}
\label{epsilonmu}
\beqa
\epsilon^{[\tau=1]}&\!\!\sim\!\!&
1+ 4\pi\, nb^3\big[j_0(k b)+j_2(k b)\big]\,,
\label{epsilon}
\\[2mm]
\mu^{[\tau=1]}&\!\!\sim\!\!&
1- 2\pi\, nb^3 \big[j_0(k b)+j_2(k b)\big]\,,
\label{mu}
\eeqa
\end{subequations}
where $n \equiv 1/l^3$ is the number  density of defects
(mean separation $l$)
and $j_p(z)$ is the spherical Bessel function of order $p$,
for example, $j_0(z)=(\sin z)/z$.
The similarity signs in Eqs.~\eqref{epsilon} and \eqref{mu}
indicate that only the $\ell=1$ multipoles have been taken into
account \cite{endnote-multipoles}.
With \eqref{disprel}, the dispersion relation is then
\begin{widetext}
\begin{equation}  
\Big(\omega^{[\tau=1]}_\gamma(k)\Big)^2 \sim
\frac{c^2\,k^2}{\big(1+4\pi\,nb^3\left[j_0(k b)+j_2(k b)\right]\big)
  \big(1-2\pi\,nb^3\left[j_0(k b)+j_2(k b)\right]\big)}\;,
\label{disp-rel-tau1}
\end{equation}
which holds for $k b \ll kl \ll 1$.
A Taylor expansion in $nb^3$ and $kb$ gives
\beqa  
\Big(\omega^{[\tau=1]}_\gamma(k)\Big)^2&\sim&
\left(1-2\pi\, nb^3 \right)\,c^2\,k^2
+ (\pi/5)\, nb^5\,c^2\,k^4+\ldots\; .
\label{disp-rel-tau1-Taylor}
\eeqa
\end{widetext}
As mentioned already, the dispersion relation \eqref{disp-rel-tau1}
has only been derived for sufficiently small values of $k$.
However, taking this expression \eqref{disp-rel-tau1} at face
value, we note that the corresponding front velocity
($v_\text{front}\equiv \lim_{k\to\infty}\,v_\text{phase}$)
would be precisely $c$.
See, e.g., Ref.~\cite{Brillouin1960} for the relevance of the front
velocity to the issues of signal propagation and causality.

For the $\tau=2$ spacetime-foam model, the dispersion relation is
found to be given by
\begin{equation}  
\Big(\omega^{[\tau=2]}_\gamma(k)\Big)^2 \sim
\frac{c^2\,k^2}{1+2\pi\,nb^3\,\big[j_0(k b)+j_2(k b)\big]}\;,
\label{disp-rel-tau2}
\end{equation}
with Taylor expansion
\beqa  
\Big(\omega^{[\tau=2]}_\gamma(k)\Big)^2 &\!\!\sim\!\!&
\left(1-2\pi\, nb^3 \right)\,c^2\,k^2
+(\pi/5)\,nb^5\,c^2\,k^4 +\ldots\,.
\label{disp-rel-tau2-Taylor}
\eeqa
Apparently, the result \eqref{disp-rel-tau2-Taylor} for randomly
orientated $\tau=2$ defects agrees, to the order shown, with the previous
result \eqref{disp-rel-tau1-Taylor} for unoriented $\tau=1$ defects.
Some  results for anisotropic defect distributions
are given in App.~\ref{sec:birefringence}.

For the $\tau=3$ spacetime-foam model, the calculation is more
complicated as the correction fields of the two ``wormhole mouths''
[spheres $S_b$ and $S_b^\prime$ in Eq.~\eqref{Mtau3}]
affect each other. Therefore, we have to work directly
with Taylor expansions in $b/d$. Giving only the leading order
terms in $b/d$, the end result is
\beqa  
\Big(\omega^{[\tau=3]}_\gamma(k)\Big)^2 &\!\!\sim\!\!&
\left(1 - (20\,\pi/3)\,nb^3\right)\,c^2\,k^2
+(2\pi/9)\,nb^3 \, d^2\,c^2\,k^4 +\ldots\,,
\label{disp-rel-tau3-Taylor}\eeqa
which holds for $k b \ll kd \ll kl \ll 1$.
For anisotropic defect distributions, some results
are again given in App.~\ref{sec:birefringence}.

In closing, it is to be emphasized that \emph{any}
localized defect (weaving error) of space responds to an
incoming electromagnetic plane wave by radiation fields
corresponding to fictitious multipoles \cite{B44}.
Only the position and relative strengths of these multipoles
depend on the detailed structure of the defect. Together with the
statistical distribution of the defects, these details then determine
the precise numerical coefficients
of the modified photon dispersion relation
written as a power series in $\vec{k}$
(see Sec.~\ref{sec:results} for further discussion).

\subsection{Scattering}
\label{sec:scattering}

In this subsection, another aspect of electromagnetic-wave
propagation is  discussed, namely, the scattering
of an incoming plane wave by $\tau=1$ defects.
Similar results are expected for  $\tau=2$ and $\tau=3^\prime$ defects.

As mentioned in Sec.~\ref{sec:photon-dispersion-relation},
the \bcs~of a $\tau=1$ defect correspond precisely
to those of a perfectly conducting sphere.
So the problem to consider is the scattering of an
electromagnetic wave by a random distribution of identical
perfectly conducting spheres with radii $b$ and mean
separation $l$, in the long-wavelength limit.
More precisely, the relevant case has $b \ll l \ll \lambda$, whereas
ideal Rayleigh scattering (incoherent scattering by randomly distributed
dipole scatterers) would have $b \ll \lambda \ll l \,$.
This means that all dipoles in a volume $k^{-3}$ radiate
\emph{coherently} and their number, $N_\text{coh} \sim (k^{-3})/(l^3)$,
appears as an extra numerical factor in the absorption coefficient
compared to standard Rayleigh scattering
(see, e.g., Sec.~I--32--5 of Ref.~\cite{F64} and
Secs.~9.6 and 9.7 of Ref.~\cite{J75}).

The relevant absorption coefficient (inverse scattering length)
is then given by
\beq
a^{[\tau=1]}_\text{scatt}
\equiv 1/L^{[\tau=1]}_\text{scatt}
       \sim \sigma_\text{dip}\; l^{-3} \; N_\text{coh} \,,
\eeq
with $\sigma_\text{dip}$ the cross section from the
electric/magnetic dipole corresponding to an individual defect,
$l^{-3}$ the number density of such dipoles (i.e., defects),\
and $N_\text{coh} \gg 1$ the coherence factor for
the $l \ll \lambda$ case discussed above.
From the calculated polarizabilities of a $\tau=1$ defect, one
has $\sigma_\text{dip} \sim k^4\,b^6$ neglecting factors of order unity.
With $N_\text{coh} \sim (k\,l)^{-3}$, the scattering length becomes
\beq\label{L-scattering}
L_\text{scatt}^{[\tau=1]} \sim k^{-1}\: \big(l/b\big)^6 \,,
\eeq
again up to factors of order unity.
Expression \eqref{L-scattering} suffices for our purpose but
can, in principle, be calculated exactly, given the statistical
distribution of defects \cite{PP62,J75}.

\subsection{Proton dispersion relation}
\label{sec:proton-dispersion-relation}

In this last subsection,
we obtain the \drs~from the Klein--Gordon
and Dirac equations for the $\tau=1$ spacetime-foam model.
For the Klein--Gordon case, similar results are expected from
the $\tau=2$ and $\tau=3^\prime$ models, but, for the Dirac case,
the expectations are less clear and a full calculation
seems to be required.

For $\tau=1$ defects and the long-wavelength approximation
$k b \ll 1$ (i.e., considering  the undisturbed harmonic fields to be
spatially constant on the scale of the defect), the heuristics
is as follows:
\begin{enumerate}
\item[(i)]
a scalar field obeying  the Klein--Gordon equation
does not require fictional sources to satisfy
the \bcs~at $|\vec x|=b$ and, therefore, the dispersion relation
is unchanged to leading order (there may,
however, be other effects such as scattering \cite{endnote-ice});
\item[(ii)]
a spinor field obeying the Dirac equation does require fictional sources
but their contributions average to zero for many randomly positioned defects
and the dispersion relation is unchanged to leading order.
\end{enumerate}

A detailed calculation (not reproduced here) gives, indeed,
unchanged constant and quadratic terms in the \dr~of a  real scalar,
at least to leading order in $k b$.
The Dirac calculation is somewhat more subtle and details are given
in App.~\ref{sec:Dirac-wave-function}.
The end result for the \dr~of a free Dirac particle,
for definiteness taken to be a proton, is:
\begin{equation}
\Big(\omega^{[\tau=1]}_p(k)\Big)^2 \sim
\hbar^{-2}\,c^4\,m_p^2 + c^2\,k^2 +  \ldots \;,
\label{disp-rel-proton-tau1}
\end{equation}
with higher-order terms neglected and  proton mass $m_p$. These
neglected higher-order terms in the proton dispersion relation
would, for example, arise from additional
factors $k^2\,b^2$ and $b^2/l^2$, resulting in possible terms
with the structure $c^2\,k^2\,(m_p^2\:c^2/\hbar^2)\,(b^4/l^2)$
and $c^2\,k^4\,(b^4/l^2)$.

The combined photon and proton dispersion-relation results will be
discussed further in the next section.

\section{Dispersion relation results} 
\label{sec:results}

\subsection{Coefficients and comments}
\label{sec:coefficients}

The different dispersion relations encountered up till now
can be summarized as follows:
\begin{widetext}
\beq
\big(\,\omega^{[\tau]}_s\,\big)^2 = \hbar^{-2}\,c^4\,m_s^2
+\left(1 + K_{\tau\,s\,2}\; b_\tau^3/l_\tau^3\,\right)\; c^2\, k^2
+K_{\tau\,s\,4}\; b_\tau^5/l_\tau^3\; c^2\, k^4 + \ldots\,,
\label{disprel-general-results}
\eeq
\end{widetext}
for defect type $\tau=1$, $2$, $3$ and particle
species $s=\gamma$, $\phi$, $p$ corresponding to the
Maxwell, Klein--Gordon, and Dirac equations, respectively.
Four technical remarks are in order.
First, the implicit assumption of \eqref{disprel-general-results}
is that the \emph{same} maximum limiting velocity $c$ holds for
all particles in the absence of defects
(that is, for particles propagating in Minkowski spacetime).
Second, the photon mass vanishes in Maxwell theory, $m_\gamma=0$,
as long as gauge invariance holds.
Third, only a few terms have been shown
explicitly in \eqref{disprel-general-results}
and, \emph{a priori}, there may be many more (even up to order $k^4$,
as explained at the end of Sec.~\ref{sec:proton-dispersion-relation}).
Fourth, a suffix $\tau$ has been added to the length scales
$b$ and $l$ of the models, since the length scale $b$ of a $\tau=1$ defect,
for example, is not the same quantity as
the length scale $b$ of a $\tau=2$ defect. But, elsewhere in the text, this
suffix is omitted, as long as it is clear which model is discussed.

In the previous section, the photon coefficients
$K_{\tau\,\gamma\,2}$ and $K_{\tau\,\gamma\,4}$ have been calculated
for all three foam models ($\tau=1$, $2$, $3$),
but those of the scalar and proton dispersion relations only
for the $\tau=1$ model. The quadratic and quartic photon coefficients
are given in Table~\ref{tab:table1}.
The quadratic proton coefficient $K_{1\,p\,2}$ from the $\tau=1$
foam model vanishes
according to Eq.~\eqref{disp-rel-proton-tau1},  as does the
scalar coefficient $K_{1\,\phi\,2}\,$.
Note that the present article considers only pointlike defects but that,
in principle, there can also be linelike and planelike defects which
give further terms in the modified \drs~\cite{endnote-linear-defects}.

\begin{table}[t]
\caption{\label{tab:table1} Quadratic and quartic coefficients $K$
in the photon dispersion relation \eqref{disprel-general-results},
for $s=\gamma$, $m_\gamma=0$, and defect type $\tau$.}
\begin{ruledtabular}
\begin{tabular}{c|cc}
  & $K_{\tau\,\gamma\,2}$
 & $K_{\tau\,\gamma\,4}$ \\
  \hline
  $\tau=1\;$  &  $-2\pi$ & $\pi/5$  \\
  $\tau=2\;$  &  $-2\pi$ & $\pi/5$  \\
  $\tau=3\;$  &  $-20\,\pi/3$ & $(2\pi/9)\;d^2/b^2$  \\
\end{tabular}
\end{ruledtabular}
\end{table}

Let us close this subsection with three general comments.
First, the modification of the quadratic coefficient of the
photon \dr, as given by Eq.~\eqref{disprel-general-results}
and Table~\ref{tab:table1}, can be of order unity
(for $b_\tau$ somewhat less than $l_\tau$) and is not suppressed by
powers of the quantum-electrodynamics coupling constant $\alpha$ or
by additional inverse powers of the large energy scale
$\Lambda \sim \hbar c/b_\tau$ (which is already impossible for
dimensional reasons, with $m_\gamma=0$ and a fixed density
factor $1/l_\tau^3\,$).
This last observation agrees with a well-known result from quantum
field theory; see, e.g., Refs.~\cite{Veltman1981,Collins-etal2004}.
Namely, if a symmetry (here, Lorentz invariance) of the quantum
field theory considered is violated by the high-energy cutoff $\Lambda$
(or by a more fundamental theory),  then, without fine tuning,
the low-energy effective theory may contain symmetry-violating terms
which are not suppressed by inverse powers of the cutoff energy
$\Lambda$.

Second, the calculated dispersion relations
\eqref{disprel-general-results}
do not contain cubic terms in $k$, consistent with general
arguments based on coordinate independence and
rotational invariance \cite{Lehnert2003}.
Furthermore, the photon dispersion relations found are the same for
both polarization modes (i.e., absence of birefringence).
For an anisotropic distribution of defects of
type $\tau=2$ or $\tau=3$, however,
the photon dispersion relations do show birefringence
but still no cubic terms;
see App.~\ref{sec:birefringence}.

Third, an important consequence of having different proton and photon
\drs~\eqref{disprel-general-results} is, as mentioned in the Introduction,
the possibility of having so-called ``vacuum Cherenkov radiation''
\cite{Beall1970,ColemanGlashow1997}. A detailed study of this process
in a somewhat different context (quantum electrodynamics with an
additional Chern--Simons term in the photonic action)
has been given in Refs.~\cite{LehnertPotting2004,KaufholdKlinkhamer2005}.

\subsection{General form}
\label{sec:general-form}

The previous results on the dispersion relation \eqref{disprel-general-results}
for the proton and photon can be combined and rewritten in the following
general form:
\begin{widetext}
\begin{subequations}
\label{disprel-general-form}
\beqa
\omega^{2}_p &\equiv& \hbar^{-2}\,
c_p^4
\,\overline{m}_p^2+
c_p^2\, k^2 + \text{O}(k^4)\,,
\label{proton-disprel-general-form}
\\[2mm]
\omega^{2}_\gamma &=&
\big(1+ \overline{\sigma}_2\, \overline{b}^3/\overline{l}^3\,\big)\; c_p^2\, k^2 +
\overline{\sigma}_4\;\overline{b}^5/\overline{l}^3\;\, c_p^2\, k^4 + \ldots\,,
\label{gamma-disprel-general-form}
\eeqa
\end{subequations}
\end{widetext}
for $0 \leq k\overline{b}, k\overline{l} \ll 1$ and sign
factors $\overline{\sigma}_2, \overline{\sigma}_4 \in \{-1,0,+1\}$.
The velocity squared $c_p^2$ is \emph{defined} as the coefficient
of the quadratic term in the proton dispersion relation
\eqref{proton-disprel-general-form} and
the effective proton mass squared $\overline{m}_p^2$
is to be identified with the experimental value.

With the results of Table~\ref{tab:table1}
and Eq.~\eqref{disp-rel-proton-tau1}, it is possible to get
the explicit expressions for the \emph{effective} length scales
$\overline{b}$ and $\overline{l}$ in terms of the \emph{fundamental}
length scales $b_\tau$ and $l_\tau$ of the spacetime model considered.
Specifically, the $\tau=1$ spacetime-foam model has
\begin{widetext}
\beq\label{eq:tau1effectivebandl}
\overline{b}
\sim
10^{-1/2}\; b_1      \,,\;\;
\overline{l}
\sim
(2\pi)^{-1/3}\; 10^{-1/2}\;l_1 \,,\;\;
\overline{\sigma}_2 =-1\,,\;\; \overline{\sigma}_4 =1 \;
\Big|^{\,[\tau = 1]}\;,
\eeq
\end{widetext}
with radius $b_1$ of the individual defects
(identical empty spheres with antipodal points identified)
and mean separation $l_1$ between the different defects.
In other words, the effective and fundamental length scales
of the $\tau=1$ model are simply proportional to each other
with coefficients of order unity.

Similar results are expected for the $\tau=2$ and $\tau=3^\prime$ models,
defined by Eqs.~\eqref{Mtau2} and \eqref{Mtau3prime}, respectively.
More generally, one could have a mixture
of different defects (types $\tau=1$, $2$, $3^\prime$, or other),
with calculable parameters $\overline{b}$, $\overline{l}$, and
$\overline{\sigma}_{2,4}$
in the photon dispersion  relation \eqref{gamma-disprel-general-form}.

For the purpose of this article, the most important result
is that the effective length scales $\overline{b}$ and $\overline{l}$
of the photon dispersion  relation \eqref{gamma-disprel-general-form}
are \emph{directly} related to the fundamental length scales of the
underlying spacetime model. This is a crucial improvement
compared to a previous calculation of anomalous effects
from a classical spacetime foam  \cite{Klinkhamer2000,KlinkhamerRuppPRD},
where the connection between effective and fundamental
length scales could not be established rigorously.

The parametrization \eqref{disprel-general-form} for the isotropic case holds true
in general and will be used in the following.
Its length scales $\overline{b}$ and $\overline{l}$ will simply be called
the \emph{average} defect size and separation, respectively.
Moreover, $\overline{b}/\overline{l}$ and $\overline{b}/\lambda$ ratios
of order one will be allowed
for, even though the calculations of Sec.~\ref{sec:photon-dispersion-relation},
leading, for example, to the identifications \eqref{eq:tau1effectivebandl},
are only valid under the technical assumptions $b/l \ll 1$ and $b/\lambda \ll 1$.
In short, the proposal is to consider a modest generalization of our
explicit results.

\section{Astrophysics bounds}
\label{sec:astrophysics-bounds}

The discussion of this section closely follows the one
of some previous articles
\cite{KlinkhamerRuppPRD,KlinkhamerRuppPRD-BR,KlinkhamerRuppNewAstRev2005},
which investigated modified dispersion relations from an entirely different
(and less general \cite{endnote-anomaly}) origin. For completeness,
we repeat the essential steps and give the original references.
Note also that, for simplicity, we focus on two particular
``gold-plated'' events but that other astrophysical input may very well
improve the bounds obtained here.

\subsection{Time-dispersion bound}
\label{sec:time-dispersion-bound}

The starting point for our first bound is the suggestion \cite{AEMNS98}
that the absence of time dispersion in a highly energetic burst of
gamma--rays can be used to obtain bounds on modified dispersion relations
(see, e.g., Refs.~\cite{Ellis-etal1999,Ellis-etal2006}
for subsequent  papers
and Ref.~\cite{Jacobson-etal2006} for a review).

 From the photon dispersion relation \eqref{gamma-disprel-general-form},
the relative change of the group velocity
$v_\text{g} \equiv \text{d}\omega/\text{d}k$
between two different wave numbers $k_1$ and $k_2$
is given by \cite{endnote-correction}:
\beqa
 \left.\frac{\Delta c}{c}\,\right|_{\, k_1,k_2} &\equiv&
 \left| \frac{v_\text{g}(k_1)-v_\text{g}(k_2)}{v_\text{g}(k_1)} \right|
\sim                          
(3/2)\, \left| k_1^2 - k_2^2 \right| \,\overline{b}^5/\overline{l}^3 \,,
\label{relativechangegroupvel}
\eeqa
where $\Delta c/c$ on the \lhs~is a convenient short-hand notation
and where $\overline{b}$ and $\overline{l}$ on the \rhs~can be interpreted
as, respectively, the average defect size and separation
(see Sec.~\ref{sec:general-form} for further discussion).

A flare of duration $\Delta t$ from an astronomical source at distance $D$,
with wave-number range
$k_1 \ll k_2 \equiv k_{\gamma,\,\text{max}} \equiv E_{\gamma,\,\text{max}}/(\hbar\, c)$,
constrains the relative change of group velocity,
$\Delta c/c \leq c\,\Delta t/D$.
Using \eqref{relativechangegroupvel}, this results in the following bound:
\begin{widetext}
\beqa
\big(\overline{b}/\overline{l}\big)^{3/2}\;\, \overline{b}
&\leq& \frac{1}{\sqrt{3/2}}\; \left(\frac{\hbar\, c}{E_{\gamma,\,\text{max}}}\right) \;
\left( \frac{c\,\Delta t}{D} \right)^{1/2}
\no\\[2mm]
& \approx &
1.2 \times 10^{-26} \:\text{m}\;
\left( \frac{2.0 \;\text{TeV}}{E_{\gamma,\,\text{max}}} \right)
\left( \frac{\Delta t}{280\;\text{s}} \right)^{1/2}
\left(  \frac{1.3 \times 10^{16}\;\text{s}}{D/c} \right)^{1/2} \,,
\label{time-disp-bound}
\eeqa
\end{widetext}
with values inserted for a $\text{TeV}$ gamma--ray flare from the
active galaxy Markarian 421 observed on May 15, 1996
at the Whipple Observatory \cite{Gaidos-etal1996,Biller-etal1999}.
(The galaxy Mkn 421 has a redshift $z\approx 0.031$ and its distance has been
taken as $D =
 c\,z/H_{0}\approx  124\:\text{Mpc}$,
for Hubble constant $H_{0}\approx 75\;\text{km}/\text{s}/\text{Mpc}$.)
The upcoming Gamma--ray Large Area Space Telescope (GLAST)
may improve bound \eqref{time-disp-bound} by a factor of $10^4$,
as discussed in App. A of Ref.~\cite{KlinkhamerRuppNewAstRev2005}.

\subsection{Scattering bound}
\label{sec:scattering-bound}

It is also possible to obtain an upper bound on the ratio
$\overline{b}/\overline{l}$ by demanding the
scattering length $L$ to be larger than the source distance $D$
or, better, larger than $D/100$ for an allowed reduction of the
intensity by a factor $F \equiv \exp(f) =\exp(100)$.
In other words, the chance for
a gamma-ray to travel over a distance $D$ would be essentially zero if
$L$ were less than $D/10^2$ (see discussion below).

The relevant expression for the scattering length $L$ from $\tau=1$
defects has been given in Sec.~\ref{sec:scattering}.
Here, we simply replace $b$ and $l$ in
result \eqref{L-scattering} by the general
parameters $\overline{b}$ and $\overline{l}$,
again allowing for the case $\overline{b}\sim\overline{l}$.
Demanding $L \gtrsim D/f$, then gives the announced bound:
\begin{widetext}
\beqa
\big(\overline{b}/\overline{l}\big)^3
&\lesssim&
\sqrt{f}\;\left( \,k_{\gamma,\,\text{max}}\, D\,\right)^{-1/2}
\no\\[2mm]
& \approx &
1.6 \times 10^{-21} \;
\left( \frac{f}{10^2} \right)^{1/2}
\left( \frac{2.0 \;\text{TeV}}{E_{\gamma,\,\text{max}}} \right)^{1/2}
\left(  \frac{3.8 \times 10^{24}\;\text{m}}{D} \right)^{1/2} \,,
\label{scattering-bound}
\eeqa
\end{widetext}
using the same notation and numerical values as in Eq.~\eqref{time-disp-bound}.

Strictly speaking, bound \eqref{scattering-bound}
is useless if $f$ is left unspecified. The problem is to decide,
given a particular source, which intensity-reduction
factor $F \equiv \exp(f)$ is needed to be absolutely sure
that its gamma-rays would not reach us if $L$ were less than $D/f$.
Practically speaking, we think that a factor $F=\exp(100)$
is already sufficient, but the reader can make up
his or her own mind. More important for bound \eqref{scattering-bound}
to make sense is that one must be certain of the source of the
observed gamma--rays and, thereby, of the distance $D$.
For the particular TeV gamma--ray flare discussed here, the identification
of the source as Mkn 421 appears to be reasonably firm \cite{Gaidos-etal1996}.

\subsection{Cherenkov bounds}
\label{sec:Cherenkov-bounds}

A further set of constraints follows from
the suggestion \cite{Beall1970,ColemanGlashow1997} that
ultra-high-energy cosmic rays (UHECRs)
can be used to search for possible Lorentz-noninvariance effects
(or possible effects from a violation of the equivalence principle).
The particular process considered here is vacuum Cherenkov radiation,
which has already been mentioned in the last paragraph
of Sec.~\ref{sec:coefficients}.

 From a highly energetic cosmic-ray event observed on
October 15, 1991 by the Fly's Eye Air Shower Detector
\cite{Bird-etal1995,Risse-etal2004},
Gagnon and Moore \cite{GagnonMoore2004}  have obtained
the  following bounds on the quadratic and quartic coefficients
of the modified photon \dr~\cite{endnote-GMbounds}:
\begin{widetext}
\begin{subequations}
\label{Cherenkov-bounds}
\beqa
-3 \times 10^{-23} &\lesssim&
\overline{\sigma}_2 \,\overline{b}^3/\overline{l}^3 \lesssim
 3 \times 10^{-23} \,,
\label{Cherenkov-bound-ratio}\\[2mm]
-\left(7 \times 10^{-39}\,\text{m}\right)^{2} &\lesssim&
\overline{\sigma}_4\,\overline{b}^5/\overline{l}^3 \lesssim
 \left(5 \times 10^{-38}\,\text{m}\right)^{2}\,,
\label{Cherenkov-bound-bsquare}
\eeqa
\end{subequations}
\end{widetext}
for length scales $\overline{b}$, $\overline{l}$
and sign factors $\overline{\sigma}_2$, $\overline{\sigma}_4$
as defined by Eqs.~\eqref{proton-disprel-general-form}
and \eqref{gamma-disprel-general-form}.
For these bounds, the primary was assumed to be a proton
with standard partonic distributions and
energy $E_p^\text{UHECR}  \approx 3 \times 10^{11} \, \text{GeV}\,$.
Note that the limiting values of bounds \eqref{Cherenkov-bound-ratio}
and \eqref{Cherenkov-bound-bsquare} scale approximately as
$\left(3 \times 10^{11} \, \text{GeV}/E_p^\text{UHECR}\right)^n$
with $n=n_a=2$ and $n=n_b=4$, respectively.
See Ref.~\cite{GagnonMoore2004} for further details  on these bounds
and App.~B of Ref.~\cite{KlinkhamerRuppNewAstRev2005}
for a heuristic discussion.

\section{Conclusion}
\label{sec:conclusion}

The time-dispersion bound \eqref{time-disp-bound} on a particular
combination of length scales from the modified photon
\dr~\eqref{gamma-disprel-general-form} is direct
and, therefore, reliable. The same holds for the scattering bound
\eqref{scattering-bound}, as long as the allowed intensity-reduction
factor is specified (see Sec.~\ref{sec:scattering-bound} for further
discussion).
The Cherenkov bounds \eqref{Cherenkov-bound-ratio}
and \eqref{Cherenkov-bound-bsquare}, however, are indirect in that they
depend on further assumptions, e.g., interactions described by
quantum electrodynamics and standard-model partonic structure
of the primary hadron. Still, the physics involved is well
understood and, therefore, also these Cherenkov bounds can be
considered to be quite reliable \cite{endnote-caveat}.

Turning to theoretical considerations, it is safe to say
that there is no real understanding of what determines the large-scale
topology of space \cite{Friedmann1924}.
With the advent of quantum theory, a similar lack of understanding
applies to the small-scale structure of space \cite{W68}.
Even so, assuming the relevance of a classical spacetime-foam model
(see discussion below),
Occam's razor suggests the model to have a \emph{single} length scale,
with average defect size $\overline{b}$
and  average defect separation $\overline{l}$ of the same order
(see Sec.~\ref{sec:general-form} for details on the
interpretation of these length scales).
Without natural explanation, it would be hard to understand
why the defect gas would be extremely rarefied,
$\overline{b}\ll\overline{l}$. In the following discussion,
we focus on the single-length-scale case but the
alternative rarefied-gas case should be kept in mind.

According to the time-dispersion bound \eqref{time-disp-bound},
a static classical spacetime foam with a single length scale
($\overline{b}\sim\overline{l}$) must have
\beqa
\overline{l}\; \Big|^\text{\,single-scale}
&\lesssim&
 10^{-26} \:\text{m}
\approx                             
\hbar  c/ \big( 2 \times 10^{10}\,\text{GeV} \big) \,,
\label{single-scale-bound}
\eeqa
which is a remarkable result compared to what can be achieved by
particle-collider experiments on Earth.
As mentioned in Sec.~\ref{sec:time-dispersion-bound},
the experimental bound \eqref{single-scale-bound} may even be improved
by a factor $10^4$ in the near future,
down to a value of the order of $10^{-30} \:\text{m}$.

But the scattering and Cherenkov bounds \eqref{scattering-bound}
and \eqref{Cherenkov-bound-ratio} lead to a much stronger conclusion:
within the validity of the model, these independent bounds
rule out a single-scale static classical spacetime foam altogether,
\beq
\overline{b}/\overline{l} \lesssim 10^{-7}\,.
\label{single-scale-ruled-out}
\eeq
The fact that a single-scale foam model is unacceptable
holds even for values of $\overline{b}\sim\overline{l}$
down to some $10^{-33}\,\text{m} \approx 10^2\times l_\text{Planck}$
(the precise definition of $l_\text{Planck}$ will be given shortly),
at which length scale a classical spacetime
may still have some relevance for describing physical
processes \cite{endnote-rarefied-gas-case}.

The unacceptability of a single-scale classical spacetime foam applies,
strictly speaking, only to the particular type of models considered.
But we do expect this conclusion to hold more generally
(recall, in particular, the remarks of the last paragraph in
Sec.~\ref{sec:photon-dispersion-relation}).
For example, also a \emph{time-dependent} classical spacetime-foam
structure with a single length scale
appears to be ruled out  \cite{endnote-timedependentdefects}.

At distances of the order of the Planck length,
$l_\text{Planck} \equiv\sqrt{G\,\hbar/c^3}
\approx 1.6 \times 10^{-35}\,\text{m}
\approx \hbar c/(1.2 \times 10^{19}\,\text{GeV})$,
it is not clear what sense to make of a classical spacetime picture
\cite{endnote-QGcalculations}. Still, at
distances of order $10^2\times l_\text{Planck}$, for example,
one does expect a classical framework to emerge and, then,
result \eqref{single-scale-ruled-out}
implies that the effective classical spacetime manifold
is remarkably smooth \cite{endnote-TeVgravity}.
If this conclusion is born out, it would suggest that either the
Planck-length fluctuations of the quantum spacetime foam
\cite{W57,W68,H78,HPP80} are somehow made inoperative over larger
distances or there is no quantum spacetime foam in the first
place \cite{endnote-effective-theory}.

\noindent\section*{ACKNOWLEDGMENTS}
The authors are grateful to  K. Busch, D. Hardtke,
C. Kaufhold, C. Rupp, and G.E. Volovik for helpful discussions.

\appendix
\section{Birefringence}
\label{sec:birefringence}

The individual defects of type $\tau=2$ and $\tau=3$ have
a preferred direction given by the unit vector $\widehat{a}$
in Figs.~\ref{fig:defecttau2} and \ref{fig:defecttau3},
respectively. An anisotropic distribution of
defects may then lead to new effects
compared to the case of isotropic distributions
considered in the main text.
This appendix presents some results for the photon
dispersion relations from  aligned $\tau=2$ and $\tau=3$ defects.

Consider, first, a highly anisotropic distribution of $\tau=2$ defects
(empty spheres with points identified by reflection in an equatorial
plane with normal vector $\widehat{a}$)
having perfect alignment of the individual $\widehat{a}$ vectors
(henceforth, indicated by a caret, $\tau=\widehat{2}\,$).
Then, the two polarization modes (denoted $\oplus$ and $\ominus$)
have dispersion relations:
\begin{widetext}
\begin{subequations}   
\begin{align}
\left(\omega^{[\tau=\widehat{2}]}_{\gamma\,\oplus}
\left(k_{||},k_{\perp}\right)\right)^2
&\sim
\frac{\big(1+6\pi\,nb\;h(kb)\;\big(k_{||}^2-2k_{\perp}^2\big)/k^4\big)\;c^2\,k^{2}}
  {\big(1+6\pi\,nb\;h(kb)/k^2\big)^2\,\big(1-12\pi\,nb\;h(kb)/k^2\big)}
  \;, \\[2mm]
\left(\omega^{[\tau=\widehat{2}]}_{\gamma\,\ominus}
\left(k_{||},k_{\perp}\right)\right)^2
&\sim
\frac{\big(1+6\pi\,nb\;h(kb)\;\big(k^2_{\perp}-2k^2_{||}\big)/k^4\big)\;c^2\,k^{2}}
  {\big(1+6\pi\,nb\;h(kb)/k^2\big)^2\,\big(1-12\pi\,nb\;h(kb)/k^2\big)}\;,
\end{align}
\end{subequations}
\end{widetext}
with an auxiliary function $h(x)\equiv \cos x - (\sin x)/x = \text{O}(x^2)$
for $x=k b \ll 1$,
parallel wave number $k_{||}\equiv |\vec k \cdot \widehat{a}|$,
perpendicular wave number
$k_{\perp}\equiv |\vec k -(\vec k \cdot \widehat{a})\,\widehat{a}|$,
and  defect number  density $n \equiv 1/l^3$.
For generic wave numbers with $k_{||}\ne k_{\perp}$,
the two polarization modes have different phase velocities
($\vec{v}_\text{phase} \equiv \widehat{k}\, \omega/|\vec{k}|\,$)
and there is birefringence.

Consider, next,
perfectly aligned $\tau=\widehat{3}$ defects, that is, wormhole-like
defects with two empty spheres identified by reflection in a
central plane (normal vector $\widehat{a}$)
and with all central planes parallel to each other
(all vectors $\widehat{a}$ aligned). In this case, we do not
have a closed expression for the dispersion relations
of the two polarization modes but rather Taylor series
(the situation considered has $k b \ll kd \ll kl \ll 1$).
Only the results for two special wave numbers are given here.
First, the photon dispersion relations for
wave number parallel to the uniform orientation $\widehat{a}$
of the defects ($k_{\perp}=0$) are equal for both polarization modes:
\begin{widetext}
\begin{align}  
\begin{split}
\left(\omega^{[\tau=\widehat{3}]}_{\gamma\,\oplus}\left(k_{||},0\right)\right)^2
=
\left(\omega^{[\tau=\widehat{3}]}_{\gamma\,\ominus}\left(k_{||},0\right)\right)^2
\sim &
\left(1-4\pi\,nb^3\right)\,c^2\,k_{||}^2 +
2\pi\,nb^3d^2\;\,c^2\,k_{||}^4+\ldots \,.
\end{split}
\end{align}
Second, the photon dispersion relations
for wave number perpendicular to
the defect orientation ($k_{||}=0$)  are different
for the two polarization modes (i.e., show birefringence):
\begin{subequations}
\begin{align}  
\begin{split}
\left(\omega^{[\tau=\widehat{3}]}_{\gamma\,\oplus}
\left(0,k_{\perp}\right)\right)^2
\sim &
\left(1+8\pi\,nb^3 \right)\,c^2\,k_{\perp}^2-
(8\pi/5)\,nb^5\,c^2\,k_{\perp}^4 +\ldots\,,
\end{split}\\[2mm]
\begin{split}
\left(\omega^{[\tau=\widehat{3}]}_{\gamma\,\ominus}
\left(0,k_{\perp}\right)\right)^2
\sim &
\left(1-16\pi\,nb^3\right)\,c^2\,k_{\perp}^2
+(16\pi/5)\,nb^5\,c^2\,k_{\perp}^4+\ldots\,,
\end{split}
\end{align}
\end{subequations}
\end{widetext}
neglecting terms suppressed by a factor of $b/d$.

The results of this appendix  make clear that birefringence only
occurs if there is some kind of ``conspiracy''
between individual asymmetric defects in the classical spacetime foam.

\section{Dirac Wave Function}
\label{sec:Dirac-wave-function}

In this Appendix, we use the Dirac representation of the
$\gamma$-matrices (for metric signature $+\,-\,-\,-\,$ and
global Minkowski coordinates)
and refer to Refs.~\cite{Rose1961,Sakurai1967,IZ80} for further details.
For simplicity, we also set $c=\hbar=1$.
The Dirac equation in Schr\"{o}dinger form reads then
\begin{equation}\label{dirac-eq}
\i\, \partial_t \, \psi(\vec{x},t)
=\big( -\i\,\vec{\alpha}\cdot\nabla + m\, \beta \,\big)\:\psi(\vec{x},t)\,,
\end{equation}
with $4 \times 4$  matrices
\begin{subequations}
\beqa\label{alphabeta-dirac-representation}
\vec{\alpha}&\equiv&\gamma_0\, \vec{\gamma}=  \left(
                  \begin{array}{cc}
                    0           \; & \;\vec{\sigma} \\
                    \vec{\sigma}\; & \;0 \\
                  \end{array}
                \right)\,,
\\[1mm]
\beta&\equiv&\gamma_0= \left(
                  \begin{array}{cc}
                    \openone_2 & 0 \\
                    0 & -\openone_2 \\
                  \end{array}
                \right)\,,
\eeqa
\end{subequations}
in terms of the $2 \times 2$ unit matrix $\openone_2$
and the $2 \times 2$ Pauli matrices $\vec{\sigma}$.

In the presence of a single $\tau=1$ defect (sphere with radius $b$
centered at $\vec{x}=0$ and antipodal identification),
we impose the following \bc~on the Dirac spinor:
\beq
\psi(-\vec{x},t) =
\i\;\widehat{x}\cdot \vec{\alpha}\;\, \psi(\vec{x},t)\;
\Big|_{\,|\vec x|=b}^{\,[\tau = 1]}\;,
\label{eq:bc-diracspinor}
\eeq
for unit vector $\widehat{x} \equiv \vec{x}/|\vec{x}|$.
[There can be an additional phase factor
$\eta\in \bC$ ($|\eta|=1$) on the \rhs~of \eqref{eq:bc-diracspinor}, which
may in principle depend on the direction, $\eta=\eta(\widehat{x})$.]
The physical motivation of \bc~\eqref{eq:bc-diracspinor} is
that the Dirac particle
moves appropriately near the defect at $|\vec{x}| = b$.
Recall that, in the first-quantized theory considered (cf. Ref.~\cite{Sakurai1967}),
the free particle is described by a wave packet which can in principle have an
arbitrarily small extension, for example, much less than $b$.
The \bc~\eqref{eq:bc-diracspinor} then makes the probability $4$--current
$j^\mu \equiv \overline{\psi} \gamma^\mu \psi=
{\psi}^\dagger \left(\openone_4,\, \vec{\alpha}\right) \psi$
well behaved near the defect at $|\vec{x}|=b\,$:
probability density  $j^0$ equal at antipodal points,
normal component of $\vec{j}$ going through,
and tangential components of $\vec{j}$ changing direction
(cf. Fig.~\ref{fig:defecttau1} and the discussion in
Sec.~\ref{sec:photon-dispersion-relation}).

The case of primary interest to us has spin--$\textstyle{\frac{1}{2}}$
particles of very high energy compared to the rest mass $m$
but wavelength still much larger than
the individual defect size $b\,$ and the mean defect separation $l$:
\beq
m \ll k \ll 1/l \ll 1/b\,.
\eeq
An appropriate initial solution of the Dirac equation over $\bR^4$
is given by
\beq\label{eq:initial-plane-wave}
\psi(\vec{x},t)_\text{in}\sim
\exp\left(\i\, k z -\i\, \omega t \right)\;
\left(\begin{array}{c}
    1 \\    0 \\    1 \\    0 \\
  \end{array}\right)  ,
\eeq
which corresponds to a positive-energy plane wave propagating
in the $z\equiv x^3$ direction.
This wave function, however, does not satisfy
the defect \bc~\eqref{eq:bc-diracspinor} for the 3--space
\eqref{Mtau1} with a single defect
centered at the point $\vec{x}=0$.

Make now the following monopole-like \emph{Ansatz}
for the required correction:
\beq\label{eq:monopole-correction}
\psi(\vec{x},t)_\text{corr}\sim
\exp\left(-\i\, \omega t \right)\;
g(r/b)\;\i\;\widehat{x}\cdot \vec{\alpha}\,
\left(\begin{array}{c}
   s_1  \\   s_2  \\   s_3  \\  s_4   \\
  \end{array}\right)  ,
\eeq
with radial coordinate $r \equiv |\vec{x}|$, normalization $g(1)=1$,
and complex constants $s_n$.
(This particular \emph{Ansatz} is motivated by the
structure of the Green's function for the Dirac operator;
cf. Sec.~34 of Ref.~\cite{Rose1961}.)
The total wave function,
\beq\label{eq:total-wave-function}
\psi(\vec{x},t)=\psi(\vec{x},t)_\text{in}+ \psi(\vec{x},t)_\text{corr} \;,
\eeq
must then satisfy the defect \bc~\eqref{eq:bc-diracspinor}
at $r=b$, in the limit $k b \to 0$.
The appropriate constant spinor $( s_1, s_2, s_3, s_4)$ in
\eqref{eq:monopole-correction} is readily found.
Also, the function $g$ in the ``near zone'' ($r \ll \lambda$) must be given by
$g(r/b)=b^2/r^2$, in order to satisfy the Dirac equation neglecting terms
of order $m b$, $k b$, and $m/k$.

All in all, we have for the corrected wave function from
a single defect centered at $\vec{x}=\vec{x}_1$:
\begin{widetext}
\beq\label{eq:wave-function-single-defect}
\psi(\vec{x},t)\sim
\exp\left(\i\, k z -\i\, \omega t \right)
\left(\begin{array}{c}
    1 \\    0 \\    1 \\    0 \\
  \end{array}\right)
-\i\, \exp\left(-\i\, \omega t \right)\;\frac{b^2}{r_1^2}\;
\left(
  \begin{array}{c}
    \cos\theta_1 \\
    \sin\theta_1 \,\exp(\i\, \phi_1)\\
    \cos\theta_1 \\
    \sin\theta_1 \,\exp(\i\, \phi_1) \\
  \end{array}
\right)\;,
\eeq
where ($r_1$,$\theta_1$,$\phi_1$) are standard spherical coordinates
with respect to the defect center $\vec{x}_1$ and the $z$ axis
from the global Minkowski coordinate system, having
 $r_1 = |\vec{x}-\vec{x}_1| \geq b$, $\,\theta_1 \in [0,\pi)$,
 and $\phi_1 \in [0,2\pi)$.

Next, sum over the contributions of $N$ identical defects
with centers $\vec{x}=\vec{x}_j$, for
$j=1, \ldots, N$.  The resulting Dirac wave function
at a point $\vec{x}$ between the defects is given by
\beq\label{eq:wave-function-many-defects}
\hspace*{-4mm}
\psi(\vec{x},t)\sim
\exp\left(\i\, k z -\i\, \omega t \right)
\left(\begin{array}{c}
    1 \\    0 \\    1 \\    0 \\
  \end{array}\right)
-\i\,\exp\left(-\i\, \omega t \right)
\,\sum_{j=1}^{N}\;  \frac{b^2}{|\vec{x}-\vec{x}_j|^2}
\left(
  \begin{array}{c}
    \cos\theta_j \\
    \sin\theta_j \,\exp(\i\, \phi_j)\\
    \cos\theta_j \\
    \sin\theta_j \,\exp(\i\, \phi_j) \\
  \end{array}
\right),
\eeq
where $\theta_j$ and $\phi_j$ are polar and azimuthal angles with
respect to the defect center $\vec{x}_j$ and the $z$ axis.
With many randomly positioned defects present ($N \gg 1$),
the entries of the second spinor
on the \rhs~of \eqref{eq:wave-function-many-defects}
average to zero and only the initial Dirac spinor remains.
\end{widetext}

For the electromagnetic case discussed
in Sec.~\ref{sec:photon-dispersion-relation},
the fictional electric/magnetic dipoles (radial dependence $\propto 1/r^3$)
are aligned by the linearly
polarized initial electric/magnetic fields \eqref{standard-plane-wave}
and there remain correction fields after averaging,
which produce a modification of the photon dispersion relation.
As mentioned above, the corrective wave function required for
the Dirac spinor is monopole-like (radial dependence $\propto 1/r^2$)
and averages to zero. This different behavior is a manifestation of
the fundamental difference between vector and spinor fields.

The final result is that the dispersion relation of a high-energy
Dirac particle is unchanged, $\omega^2   \sim k^2$,
at least up to leading order in $m/k$, $k l$, and $b/l$.
For an initial spinor wave function at rest
($k=0$ and $b \ll l \ll 1/m$),
a similar calculation gives $\omega^2 \sim m^2$.
(Note that this is an independent calculation as Lorentz
invariance does not hold.)
Combined, the \dr~of a free Dirac particle
(for definiteness, taken to be a proton) reads
\begin{equation}\label{eq:disprel-diracspinor}
\Big(\,\omega^{[\tau=1]}_p\,\Big)^2 \sim m^2 +k^2+\ldots\;,
\end{equation}
to leading order in $m l$, $k l$, and $b/l$.

To summarize, we have found in this appendix that
the classical spacetime-foam model considered affects the
quadratic coefficient of the proton dispersion relation
differently than the one of the photon dispersion relation.
However, the calculation
performed here was only in the context of the first-quantized theory
and a proper second-quantized calculation is left for the future.



\begin{thebibliography}{99}
\bibitem{endnote-multipoles}
Higher-order multipoles ($\ell$ $=$ $3, 5, 7, \ldots$)
contribute to the effective permeabilities with a factor
$b^{\ell+2}\left[j_{\ell-1}(k b)+j_{\ell+1}(k b)\right]$
and are, therefore, suppressed by additional powers of $b$.
Note the different letter type used for multipole $\ell$
and defect separation $l$.
\bibitem{endnote-correction}
The numerical factor $2$ on the \rhs~of Eq.~(6.3) in
Ref.~\cite{KlinkhamerRuppPRD} must be replaced by $3/2$.
\bibitem{endnote-caveat}
The only \emph{caveat} we have regards the calculation of
the proton \dr~and is stated in the last sentence
of App.~\ref{sec:Dirac-wave-function}.
For this reason, we have employed the parametrization
\eqref{disprel-general-form}, which is valid generally,
even though the interpretation of the length scales $\overline{b}$
and $\overline{l}$ may be less transparent than in,
for example, Eq.~\eqref{eq:tau1effectivebandl}.
\bibitem{endnote-rarefied-gas-case}
For completeness, we mention the rarefied-gas interpretation
of our results. If somehow the ratio $\overline{b}/\overline{l}$
is very small and saturates bound \eqref{Cherenkov-bound-ratio},
$\overline{b}/\overline{l} \sim 3 \times 10^{-8}$,
then bound \eqref{Cherenkov-bound-bsquare} gives
$\overline{b} < 9 \times 10^{-27}\,\text{m}$.
Hence, the very small lengths ($10^{-38}\,\text{m}$)
appearing in the limits of bound \eqref{Cherenkov-bound-bsquare}
need not imply equally small fundamental length scales
if there is a small parameter in the game
(here, $\overline{b}/\overline{l}$). Note also that,
for the models considered in Sec.~\ref{sec:defect-types},
a perfectly smooth 3--space is recovered in the
limit $l\to \infty$, for a fixed value of $b$
(and a fixed value of $d$ in the $\tau=3$ model).
\bibitem{endnote-timedependentdefects}
As a concrete example, take the $\tau = 1$ defects of
Sec.~\ref{sec:defect-types} to be time-dependent,
with typical lifetime $\delta t$ and average re-appearance time
$\Delta t$.
In other words, a single defect with fixed position in space
is ``on'' for a fraction $\delta t/\Delta t$
of the time and the different defects
occupy  a fraction $n b^3 \equiv b^3/l^3$ of space.
(The implicit assumption here is that topology change is allowed;
see, e.g., Chap. 6 of Ref.~\cite{V96} for further discussion.
Alternatively, one might consider nonvanishing time-dependent
defect radii $b(t)$ with $b(t)=b$ for the time intervals $\delta t$
and $0 < b(t) \ll b$ otherwise.)
In a long-wavelength calculation along the lines of
Sec.~\ref{sec:photon-dispersion-relation}, there are then
defects hopping around in the nearly constant electric and
magnetic fields of the initial plane wave.
For $\delta t \leq \Delta t \ll 2\pi/\omega$ and $b \ll   l \ll \lambda\,$,
the fictional multipoles turn on or off rapidly
and one naively expects effective permeabilities
$\epsilon \sim 1+ 4\pi F$ and $\mu \sim 1-2\pi F$,
with an excluded-spacetime-volume
factor $F \sim (b^3/l^3)\times(\delta t/\Delta t)$.
The modified photon dispersion relation is now
$\omega^2\sim \left(1-2\pi F \right)\,c^2\,k^2 +\text{O}(k^4)$.
Taking the quadratic coefficient of the proton
\dr~to be unchanged, an astrophysics bound similar to
\eqref{Cherenkov-bound-ratio} applies to $-2\pi F$
and apparently rules out such a
single-scale spacetime foam with $b \sim  l \sim c\delta t \sim c\Delta t$.
But this tentative conclusion
remains to be confirmed by a detailed calculation.
\end{thebibliography}
\end{document}